\documentclass[10pt,aps,twocolumn,showpacs,pra]{revtex4}
\usepackage{amssymb,amsfonts,amsmath,wrapfig,mathrsfs,mathbbol}
\usepackage{graphicx}

\begin{document}

\title{Coherent Control of Vibrational State Population in a Nonpolar Molecule}

\author{A. Pic\'{o}n$^{1}$, J. Biegert$^{2}$, A. Jaron-Becker$^{1}$, and A. Becker$^{1}$}

\affiliation{$^{1}$JILA and Department of Physics, University of Colorado at Boulder, Boulder, Colorado 80309-0440, USA}

\affiliation{$^{2}$ICFO, Institut de Ciencies Fotoniques, Mediterranean Technology Park, E-08860 Castelldefels (Barcelona), Spain and ICREA - Instituci\'o Catalana de Recerca i Estudis Avan\c cats, E-08010 Barcelona, Spain}

\date{\today}

\begin{abstract}
A coherent control scheme for the population distribution in the vibrational states of nonpolar molecules is proposed.
Our theoretical analysis and results of numerical simulations for the interaction of the hydrogen molecular ion in its
electronic ground state with an infrared laser pulse reveal a selective two-photon transition between the vibrational states
via a coupling with the first excited dissociative state. We demonstrate that for a given temporal intensity profile the population transfer between vibrational states,
or a superposition of vibrational states, can be made complete for a single chirped pulse or a train of chirped pulses,
which accounts for the accumulated phase difference due to the AC Stark effect. Effects of a spatial intensity
(or, focal) averaging are discussed.
\end{abstract}

\maketitle

\section{Introduction}

Control of a molecular process and, in particular, control of the quantum state of a molecule
is a long-standing goal in molecular physics, chemical physics, chemistry, and, more
recently, strong-field
physics and quantum information technology.
Light fields generated from femtosecond laser pulses have been tailored to manipulate
the population in a molecule to a specific target state. Applications range from the control of
the motion and shape of a wavefunction \cite{weinacht99} to the control over
molecular fragments in a simple reaction \cite{assion98}. Moreover, novel quantum
technologies such as ultracold chemical reactions \cite{krems08}, frequency metrology
of fundamental constants \cite{koelemeij07} or
molecular quantum bits for quantum information processing \cite{demille02} require
the control over the internal quantum states of a molecule.
In many of these applications it is desirable to
completely evolve the molecular system in a robust way to the desired target quantum state.

Much progress in the preparation and manipulation of the quantum state of neutral diatomic polar molecules \cite{viteau08,ni08,deiglmayer08,lang08,ospelkaus10,danzl10}
and molecular ions \cite{staanum10,schneider10} has been achieved recently. Several experimental approaches,
e.g. photoassociation of ultracold atoms involving Feshbach resonances, spontaneous decay and optical schemes
such as stimulated Raman adiabatic passage (STIRAP), have been developed. Here, we present a theoretical study and
results of numerical simulations exploring an alternative method to manipulate and control the vibrational state population in a molecule.
We performed our analysis for
the simplest molecular ion, namely the hydrogen molecular ion. However, our results likely hold for other molecules as well.
Since H$_2^+$ is a nonpolar molecular ion, optical dipole transitions between vibrational states are forbidden.
Furthermore, the first excited electronic state ($2p \sigma_u$) is a dissociative state and higher electronic states are rather weakly bound.
Thus, control and manipulation of the internal quantum state of H$_2^+$ pose a challenge, since the application of
methods such as STIRAP is limited due to the reduction in efficiency for a transition via continuum states, e.g. molecular dissociative states \cite{peters05}.

\begin{figure}
\centering\includegraphics[width=8cm]{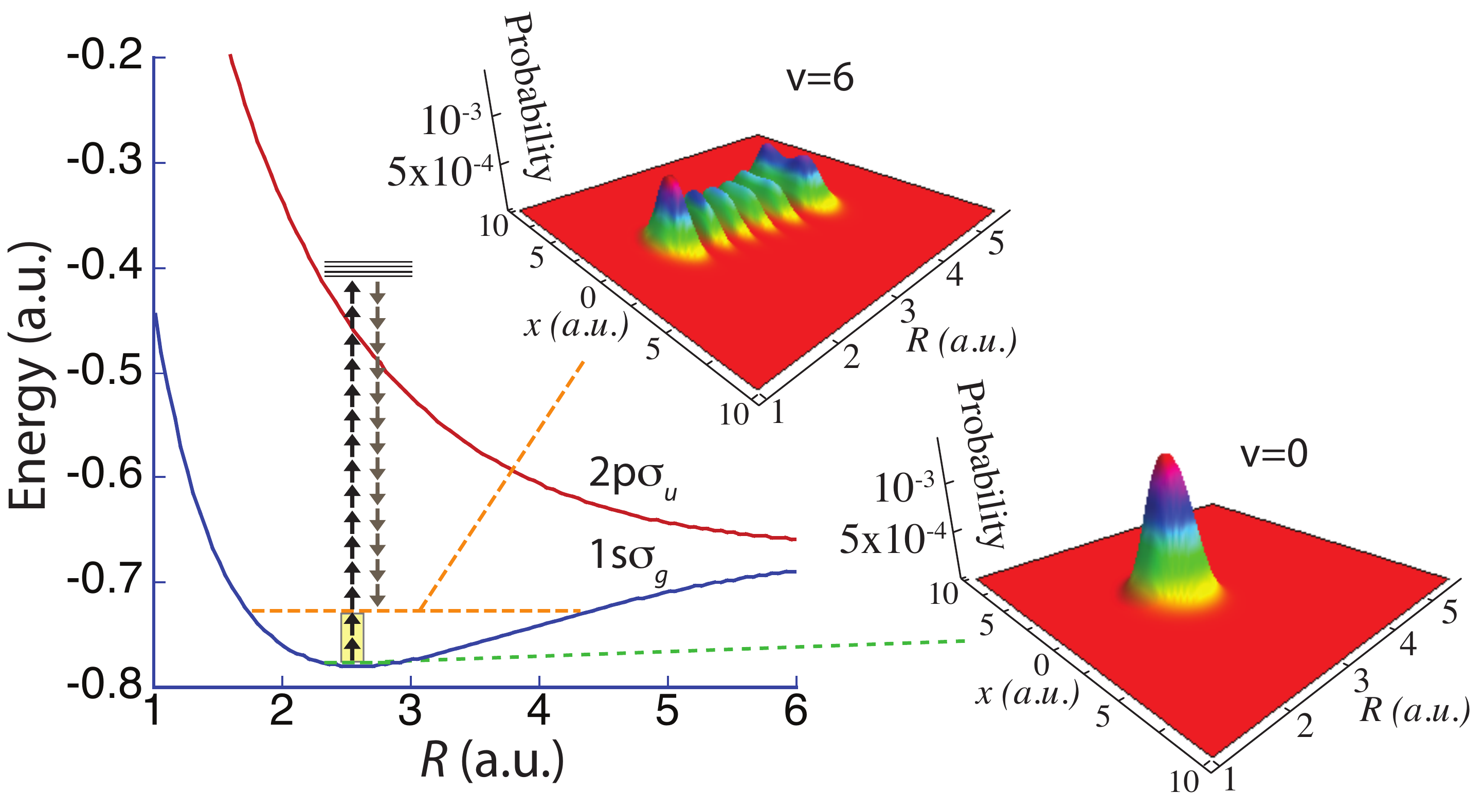}
\caption{
(Color online) Coherent control scheme for population transfer between vibrational states
of a nonpolar molecule such as H$_2^+$.
Vibrational levels (here, $\nu=0$ and $\nu=6$) of the ground electronic state ($1s\sigma_g$) of H$_2^+$ are coupled via multiphoton transitions with the first excited dissociative state ($2p\sigma_u$). Efficient control over the population transfer between the states can be achieved for near-resonant net two-photon transitions.
}
\label{fig:Energy_Levels}
\end{figure}

A sketch of the control scheme is shown in Fig. \ref{fig:Energy_Levels}. We
consider to transfer the population between two vibrational states
($\nu=0$ and $\nu=6$, as an example in Fig. \ref{fig:Energy_Levels})
in the electronic ground state of H$_2^+$ ($1s\sigma_g$).
Both states are coupled by an intense infrared laser pulse via multiphoton transitions
with the first excited dissociative state ($2p \sigma_u$). As we will show below,
population is transfered from one vibrational state (or, a superposition of
states) to the other if the photon energy approximately equals half of the
energy difference between the two vibrational states.
The transfer can be made complete
by application of a chirped laser pulse or via coherent accumulation using a train
of chirped pulses, which take account of the presence of the
dynamic Stark shifts.

The present scheme resembles analogy with well-known processes such as STIRAP \cite{bergmann98},
laser-induced continuum structure (LICS) \cite{Knight_Laser_1990} or Raman chirped adiabatic passage (RCAP)
\cite{Chelkowski_Raman_1997} but it also differs in a number of characteristic features. For example, the energy difference between
the excited state and the initial or final states is much larger than the photon energy and, consequently,
the transition via the excited state is a {\it multiphoton} transition. Traditionally, these kind of coherent control schemes are based on one-photon
transitions between the states involved. Related to this difference, application of the present scheme
is based on the interaction with just {\it one} laser pulse, while often two laser pulses operating at
different frequencies are used. As in the previous schemes, maximum population transfer is achieved for
a certain laser frequency. Interestingly however, the related condition corresponds to a two-photon resonance between
the initial and the final state.

The rest of the paper is organized as follows: We first present in Section II.A. the theoretical model for the hydrogen molecular ion, which
we used for the numerical simulations of the population transfer between two vibrational states. Based on the numerical results we
will identify the condition needed to maximize the population transfer in the field of an unchirped pulse, namely a slight detuning
from the two-photon resonance between the initial and final states. Next, in section II.B., we analyze the process using a simplified model which takes account of
just two electronic states. The analysis shows that the transfer can be made complete using a chirped pulse and an analytical formula for
the chirp will be derived. In section II.C. we verify our analytical predictions by performing calculations using the two-state model and the full
numerical simulations for transfers between individual states as well as superposition of states. Results of both set of calculations
are in perfect agreement showing the reliability for our analysis based on the two-state model.
Finally, we investigate the effect of a spatial intensity (or, focal) averaging before we conclude with the main findings presented in this paper.

\section{Theoretical models and results}

We analyzed the control scheme using analytical calculations as well as numerical simulations of the time-dependent Schr\"odinger equation (TDSE).
We first present the results of our numerical simulations for interaction of the molecular ion with an unchirped pulse at wavelengths in the infrared.
This will lead us to an analysis of the population transfer within a two-state model. Based on the conclusions of our model analysis we will then verify
by performing calculations using both the two-state model as well as the full ab-initio numerical simulations that the transfer can be made complete for a chirped pulse.

\subsection{Two-photon resonance condition}


\begin{table}[t]
\begin{center}
\begin{tabular}{|c|c|}
\hline
Vibrational State & Energy (a.u.)\\
\hline
$\nu=0$ & -0.776\\
\hline
$\nu=1$ & -0.767\\
\hline
$\nu=2$ & -0.758\\
\hline
$\nu=3$ & -0.749\\
\hline
$\nu=4$ & -0.741\\
\hline
\end{tabular}
\caption{\label{energies}
Energies of the first vibrational levels in the ground electronic state of the H$_2^+$ model used in the present study.
}
\end{center}
\end{table}

For the numerical simulations we used a two-dimensional model of the hydrogen molecular ion, in which the motion of the electron as well as the motion of the protons are restricted along the polarization direction of the external field. The Hamiltonian of the system is given in length gauge as (Hartree atomic units, $e=m=\hbar=1$, are used throughout) \cite{kulander96}:
\begin{eqnarray} \label{Hamiltonian}
H(x,R,t) &=&
\frac{P^{2}}{2\mu_{p}} + \frac{p^2}{2}
+ x {\cal E}(t) + \frac{1}{\sqrt{R^{2}+\alpha_{p}}}
\\
&&
- \frac{1}{\sqrt{\left(x-\frac{R}{2}\right)^{2} + \alpha_{e}}}
- \frac{1}{\sqrt{\left(x+\frac{R}{2}\right)^{2} + \alpha_{e}}} \; ,
\nonumber
\end{eqnarray}
where ($P$, $R$), and ($p$, $z$) are momentum operators and positions of the relative coordinate of the two protons and the electron, respectively. $\mu_p = M/2$ is the reduced mass with $M = 1836$ is the mass of the proton, and $\alpha_e = 1$ and $\alpha_p = 0.03$ are soft-core parameters \cite{kulander96}. ${\cal E}(t) = {\cal E}_0\cos(\omega t)\sin^2(\pi t/NT)$ is the electric field of the linearly polarized laser pulse. The corresponding TDSE was solved using the Crank-Nicolson algorithm on a grid with spatial spacings of $\Delta R$=0.03 au and $\Delta z$=0.1 au, the time step was less than $\Delta t$= 0.05 au. The vibrational levels in the ground electronic state of the model were found by imaginary time propagation and the corresponding energies are given in Table 1.

\begin{figure}
\centering\includegraphics[width=7cm]{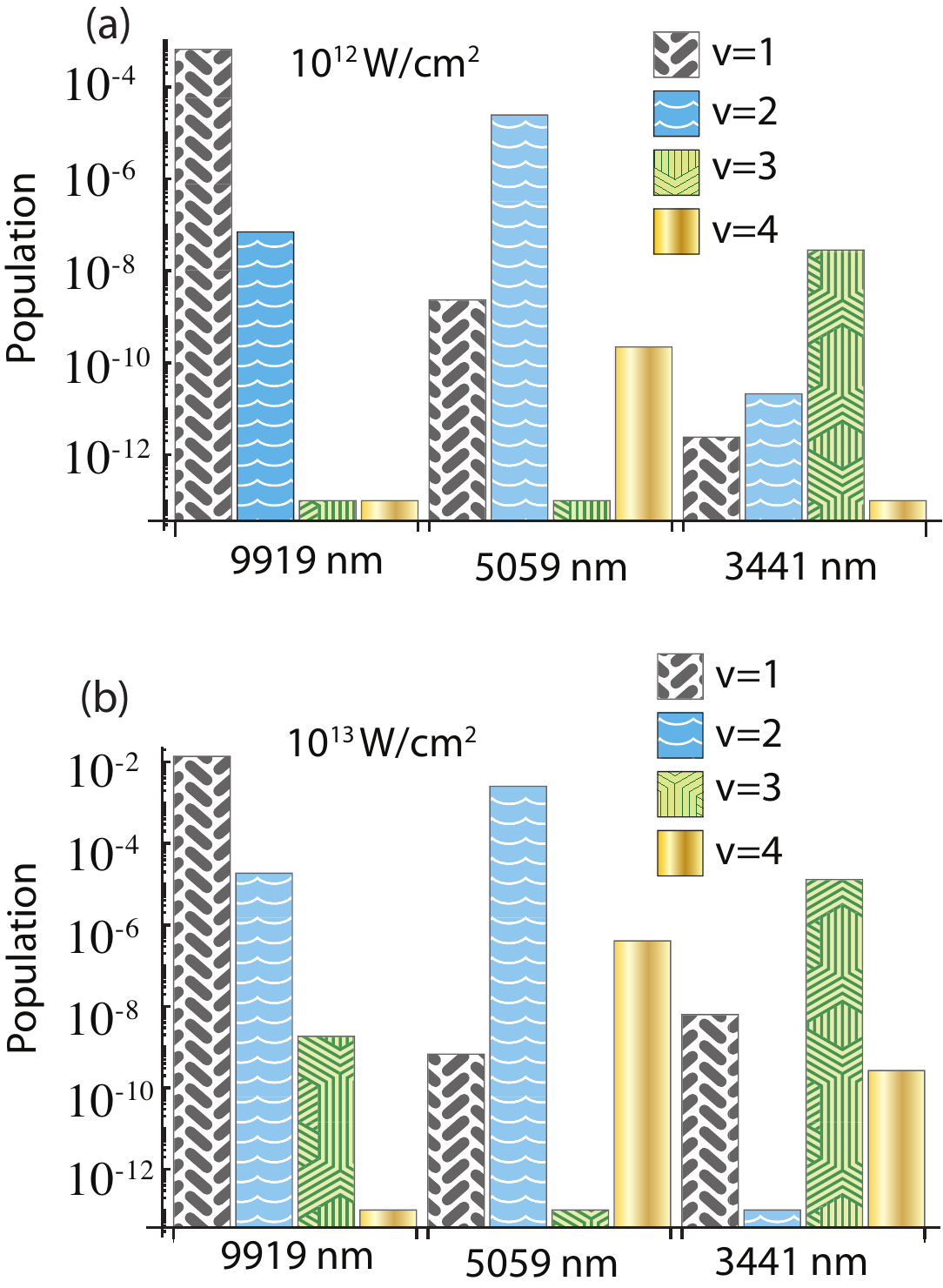}
\caption{
(Color online) Final population of the first vibrational states by interaction with laser pulses at 9919.9 nm, 5059.3 nm and 3441.0 nm, respectively.
Results of the projections to $\nu=5$ and $\nu=6$ are below $10^{-13}$ and not shown. The number of cycles was 10 in each simulation, and the peak intensity was (a) $10^{12}$ W/cm$^2$, and (b) $10^{13}$ W/cm$^2$.
}
\label{fig:Selective_Excitation}
\end{figure}

First, we performed a series of calculations in which we prepared the H$_2^+$ model system in the ground vibrational state. We tried to find the optimal wavelength of the light field for a selective population transfer into one of the excited vibrational states. In Fig. \ref{fig:Selective_Excitation} we present our results for the final populations in the first four excited states driven by laser pulses at $\lambda$ = 9919.9 nm, $\lambda$ = 5059.3 nm and $\lambda$ = 3441.0 nm. The photon energy is equal to half of the energy difference between the first, second, and third excited state and the ground state, respectively. The pulse length was 10 field cycles and the peak intensity was (a) 10$^{12}$ W/cm$^2$, and (b) 10$^{13}$ W/cm$^2$.

The results show that a particular vibrational state can be selectively excited via a (net) two-photon transition from the initially prepared ground state. The populations in all the other states are at least three orders of magnitude smaller as compared to the maximum excited state population. However, we observe a transfer of population of about 1\% only, or even less. We note that for long pulses we observe Rabi-like oscillations but without surpassing the 10\% level for the population in the excited state.

\begin{figure}
\centering\includegraphics[width=7cm]{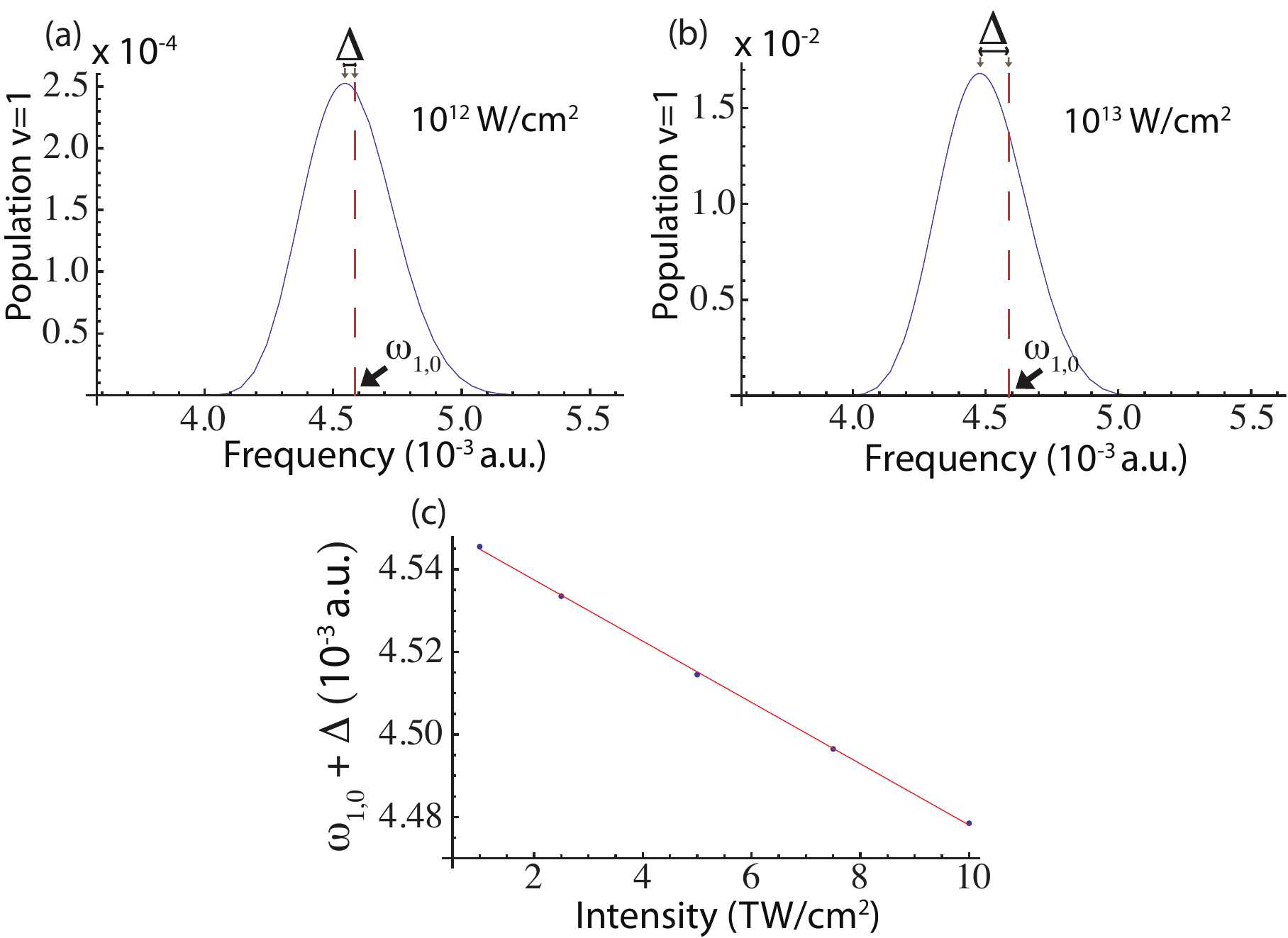}
\caption{
(Color online) Final population of $\nu=1$-state as a function of the laser frequency.  Note that the maximum occurs for a slight detuning $\Delta$ from the two-photon resonance frequency $\omega_{1,0} = (E_{\nu=1}-E_{\nu=0})/2$ au. The number of cycles was 10 in each simulation and the peak intensity was (a) $10^{12}$ W/cm$^2$ and (b) $10^{13}$ W/cm$^2$. (c) Detuning $\Delta$ as a function of intensity.
}
\label{fig:Detuning_9919}
\end{figure}
\begin{figure}
\centering\includegraphics[width=7cm]{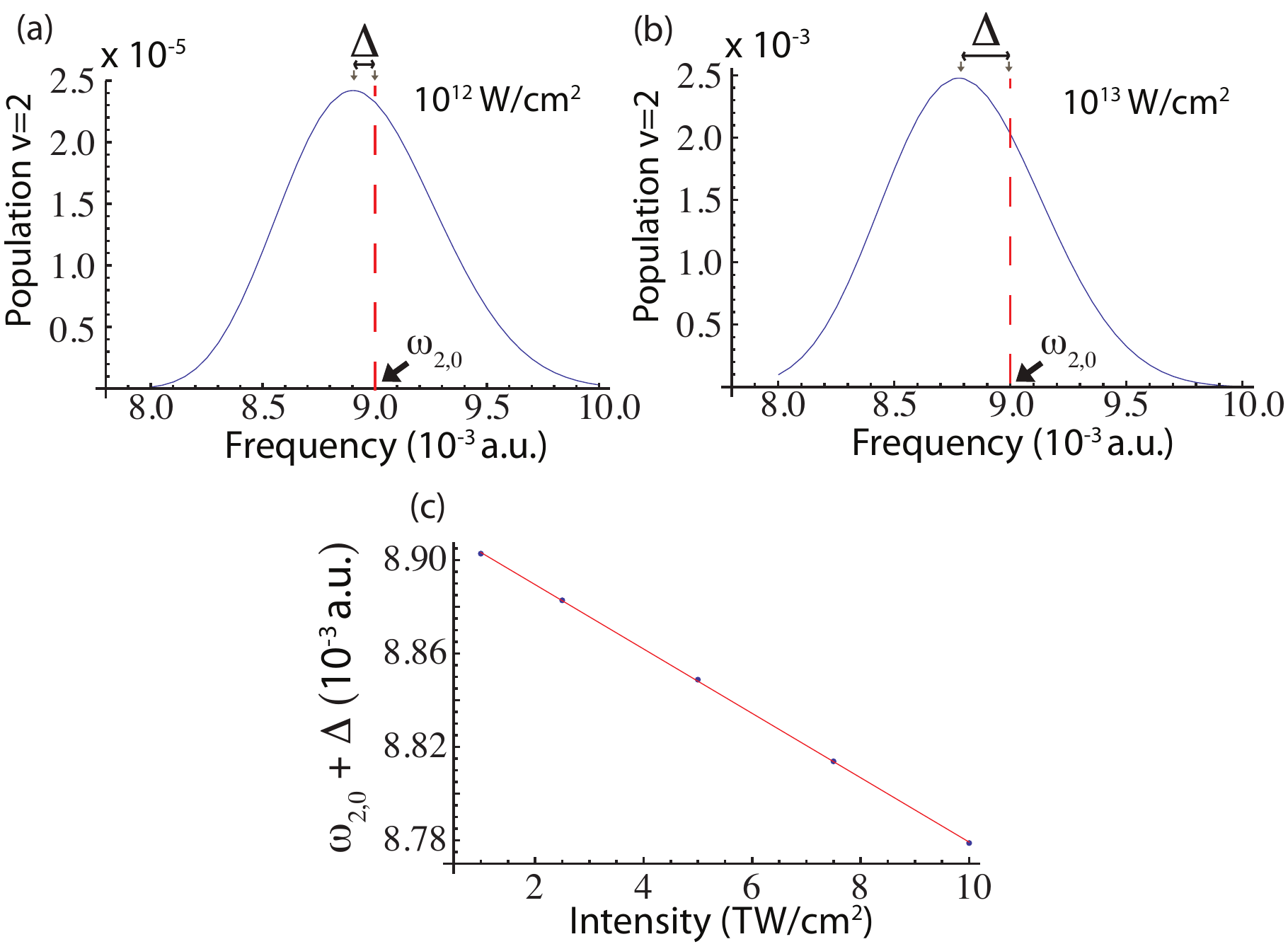}
\caption{
(Color online) Same as Fig.\ \ref{fig:Detuning_9919}, but for $\nu=2$-state.
}
\label{fig:Detuning_5059}
\end{figure}

{Next, we present in Fig. \ref{fig:Detuning_9919} results for the population in the $\nu=1$ state as a function of the laser wavelength for two different peak intensities, (a) $10^{12}$ W/cm$^2$ and (b) $10^{13}$ W/cm$^2$. The number of cycles is kept the same as in Fig. \ref{fig:Selective_Excitation}.
The results show that the population transfer is maximum for a slight detuning $\Delta$ from the two-photon resonance condition $\omega_{1,0} = (E_{\nu=1}-E_{\nu=0})/2$.
The detuning is found to depend linearly on the intensity (see Fig. \ref{fig:Detuning_9919}(c)).
These findings appear to hold in general independent of the final state, since we find the same linear dependence for excitation of the $\nu=2$ state (Fig. \ref{fig:Detuning_5059}).

\subsection{Analytical model}

Based on the findings from our numerical simulations we would like to address the following questions: What is the physical mechanism of this population transfer between vibrational states in a nonpolar molecule?
What are the origins of the near-resonant two-photon condition and the detuning? And, finally, is it possible to enhance the efficiency of the population transfer? In order to answer these questions, we developed the following analytical model.

We assume that the photon energy is small as compared to the energy difference between the electronic states and the peak intensity is sufficiently small such that the state of the molecular ion,
which is initially prepared in a superposition of vibrational states of the electronic ground state ($1s\sigma_g$), is well described
by a superposition of the two lowest electronic states at any time $t$ during the interaction (in Born-Oppenheimer approximation):
\begin{eqnarray} \label{wavefunction}
\Psi(x,R,t)
&=&
\phi_{\sigma_g}(x,R) \sum_{\nu} a_{\nu}(t) \,\psi_{\nu}(R)
\nonumber
\\
&&
+ \phi_{\sigma_u}(x,R)\int dk \, b_{k}(t) \,\psi_{k}(R) \; .
\end{eqnarray}
This reduction to a two-level model is justified within the range of laser parameters considered here, since we do not observe any ionization, dissociation or population to higher excited states in our full numerical simulations. As we will show in section II.C below, the results obtained using the two-level model do agree very well with those of the full numerical simulations.

Inserting Eq. (\ref{wavefunction}) in the Schr\"odinger equation
$i \partial \Psi(x,R,t) /\partial t = H(x,R,t) \Psi(x,R,t)$, multiplying by $\phi_{\sigma_g}^{*}(x,R)\psi_{\nu}^{*}(R)$ or $\phi_{\sigma_u}^{*}(x,R)\psi_{k}^{*}(R) $ from the left
and integrating with respect to $x$ and $R$ we obtain a set of equations for the amplitudes $a_{\nu}(t)$ and $b_k(t)$:
\begin{eqnarray}
i \; \dot{a}_{\nu}(t)
&=&
E_{\sigma_g,\nu} \,a_{\nu}(t)
+ {\cal E}(t) \int dk\; \mu_{\nu,k}\, b_k(t)  \, , \label{Schrodinger_VS_1}
\\
i \; \dot{b}_{k} (t)
&=&
E_{\sigma_u,k} \,b_k(t)
+ {\cal E}(t) \sum_{\nu}\,  \mu_{\nu,k}^{*}\, a_{\nu}(t) \, , \label{Schrodinger_VS_2}
\end{eqnarray}
with $\mu_{\nu,k} \equiv \langle\phi_{\sigma_g}(x,R)\psi_{\nu}(R) \vert x \vert \phi_{\sigma_u}(x,R)\psi_{k}(R)\rangle$, where the brackets indicate integrations over $x$ and $R$. $E_{\sigma_g,\nu}$ and $E_{\sigma_u,k}$ are the field-free energies of the states. Eqs. (\ref{Schrodinger_VS_1}) and (\ref{Schrodinger_VS_2}) represent the coupling of the vibrational states in the electronic ground state with the dissociative state. Since we are interested in the evolution of the population in the vibrational states, we integrate Eq. (\ref{Schrodinger_VS_2}) with respect to time and then substitute the result into Eq. (\ref{Schrodinger_VS_1}) to get:
\begin{eqnarray}
\dot{c}_{\nu}(t) &=&
- \sum_{\nu'} \int dk\; \mu_{\nu',k}^{*} \,\mu_{\nu,k} \times
\nonumber
\\
&&
\int_{0}^{t} dt'
{\cal E}(t) {\cal E}(t')
c_{\nu'}(t')
e^{i\Delta E_{\nu',k} t'}
e^{-i\Delta E_{\nu,k} t} \; , \label{Formula_No_Depletion_1}
\end{eqnarray}
where $c_{\nu}(t) = a_{v}(t)e^{iE_{\sigma_g,\nu}t}$ with $\vert a_{v}(t)\vert=\vert c_{v}(t)\vert$
and $\Delta E_{v,k} = E_{\sigma_u,k} - E_{\sigma_g,\nu}$.
If the laser frequency is far off-resonant, i.e. $\Delta E_{v,k}\gg \omega$,
we can assume that
the electric field strength ${\cal E}(t)$ and $c_{\nu}(t)$ are slowly varying in time compared to the exponential term, which depends on the energy difference $\Delta E_{\nu,k}$,
and perform the integral in Eq. (\ref{Formula_No_Depletion_1}) approximately to get:
\begin{eqnarray} \label{Formula_No_Depletion_Quick_Approx}
i\; \dot{c}_{\nu}(t) = - \sum_{\nu'} \mu_{\nu,\nu'}^{2} {\cal E}^{2}(t)
c_{\nu'}(t)\; e^{-i (E_{\nu'}-E_{\nu}) t}\; ,
\end{eqnarray}
where $ \mu_{\nu,\nu'}^{2} \equiv \int dk\; \mu_{\nu',k}^{*} \,\mu_{\nu,k}/\Delta E_{\nu',k}$. For the sake of simplicity we defined $E_{\sigma_g,\nu}\equiv E_{\nu}$. The temporal evolution of the amplitude $c_{\nu}$
depends quadratically on the electric field strengths, which gives rise to the two-photon transition between the vibrational states
observed in the numerical simulations.

Please note that the current control mechanism is based on {\it multiphoton} transitions for which the off-resonant condition $\Delta E_{v,k}\gg \omega$ holds.
This is the main difference between the present mechanism to induce transitions between different internal states of the molecule and processes
such as LICS (laser-induced continuum structure, \cite{Knight_Laser_1990}) and STIRAP \cite{bergmann98}.
For example, in STIRAP the off-resonant condition $\Delta E_{v,k}\gg \omega$ does not hold and Eq. (\ref{Formula_No_Depletion_1}) cannot be further simplified as in the present
analysis.

Using the two-level approximation between an initial ($\nu_i$) and a final ($\nu_f$) vibrational state
we obtain:
\begin{eqnarray}
\lefteqn{i \left(
\begin{array}{c}
\dot{c}_{\nu_i}
\\
\dot{c}_{\nu_f}
\end{array} \right)
=}
\nonumber
\\
&&
- {\cal E}^{2}(t)
\left( \begin{array}{cc}
\mu_{\nu_i,\nu_i}^{2} & \mu_{\nu_i,\nu_f}^{2} \,e^{-i \Delta E t} \\
\mu_{\nu_f,\nu_i}^{2} \,e^{i \Delta E t}  & \mu_{\nu_f,\nu_f}^{2} \\
\end{array} \right)\left(
\begin{array}{c}
c_{\nu_i}
\\
c_{\nu_f}
\end{array} \right) \; . \label{2Level_Matrix_VS}
\end{eqnarray}
where $\Delta E \equiv E_{\nu_f}-E_{\nu_i}$. Note that
$\mu_{\nu_i,\nu_f}^{2}\neq\mu_{\nu_f,\nu_i}^{*2}$, if $\Delta E_{\nu_i,k}\neq \Delta E_{\nu_f,k}$.
In general, this gives rise to a non-Hermitian Hamiltonian in Eq. (\ref{2Level_Matrix_VS}) and, hence, a
loss of population in the two-level system via excitation into the dissociative state.
The loss becomes negligible, if $\Delta E \ll \Delta E_{\nu_i,k} \approx \Delta E_{\nu_f,k}$.
This is the case for the (low) vibrational states in H$_2^+$, correspondingly we did not observe any dissociation
of H$_2^+$ in our numerical simulations.

Transforming $c_{\nu_j} (t) = \exp(i\int_{t_{0}}^{t} {\cal E}^{2}(t')\mu_{\nu_j,\nu_j}^{2} dt') \tilde{c}_{\nu_j}(t)$
(with $j = i, f$) yields:
\begin{eqnarray}
i
\left(
\begin{array}{c}
\dot{\tilde{c}}_{\nu_i} (t)
\\
\dot{\tilde{c}}_{\nu_f} (t)
\end{array}
\right)
=
\left( \begin{array}{cc}
0  &  \sigma_{\nu_i,\nu_f}
\\
\sigma_{\nu_f,\nu_i} & 0
\\
\end{array} \right)
\left(
\begin{array}{c}
\tilde{c}_{\nu_i} (t) \\ \tilde{c}_{\nu_f} (t)
\end{array} \right) \; , \label{2Level_Matrix_VS_free}
\end{eqnarray}
with
\begin{equation}
\sigma_{\nu_i,\nu_f}
= -\mu_{\nu_i,\nu_f}^{2} {\cal E}^{2}(t) \,e^{-i\left( \Delta E t + [\mu_{\nu_i,\nu_i}^{2}-\mu_{\nu_f,\nu_f}^{2}] \int_{t_{0}}^{t} {\cal E}^{2} (t') dt'\right)} \; .
\label{final}
\end{equation}
$\mu_{\nu_j,\nu_j}^{2}$ (with $j=i, f$) represents the time-dependent Stark shift of the vibrational state $j$, which results in a phase shift between the states during the interaction with the field. A similar result was obtained in Ref. \cite{Trallero_Coherent_2005,Trallero_Strong_2006} for transitions between atomic bound electronic states. Please also note that for $\mu_{\nu_i,\nu_i}^{2}=\mu_{\nu_f,\nu_f}^{2}$ Eq. (\ref{final}) reduces to the standard two-level system coupled via a two-photon transitions with an electric field \cite{Eberly_Optical_1987}.

According to the analysis above we expect that a complete coherent transfer of the population from one vibrational state to the other can be achieved by
designing a chirped pulse in which the frequency variation follows the time-dependent accumulation of the Stark phase shift as:
\begin{eqnarray}
2 \omega(t)\,t
&=&
\Delta E \, t + \left[ \mu_{\nu_i,\nu_i}^{2}-\mu_{\nu_f,\nu_f}^{2} \right]
\int_{t_{0}}^{t} {\cal E}^{2} (t') dt' \nonumber
\\
&=&
2 (\omega_{f,i} \,t + \delta (t)) \; , \label{Chirp_Condition}
\end{eqnarray}
where $\omega_{f,i}$ satisfies the two-photon resonance condition.

In general $\mu_{v,v}^{2}-\mu_{v',v'}^{2}$ is unknown, in our numerical model as well as in an experiment. However, by using a chirped pulse with $\delta (t) = a \int_{t_{0}}^{t} {\cal E}^{2} (t') dt'$ and sweeping the constant
$a$ one can optimize the transition probability. We may note parenthetically that this method may be also applied to gain information about an unknown molecular system, which we however will not further explore here.

\subsection{Population transfer with chirped pulses}

\begin{figure}
\centering\includegraphics[width=8cm]{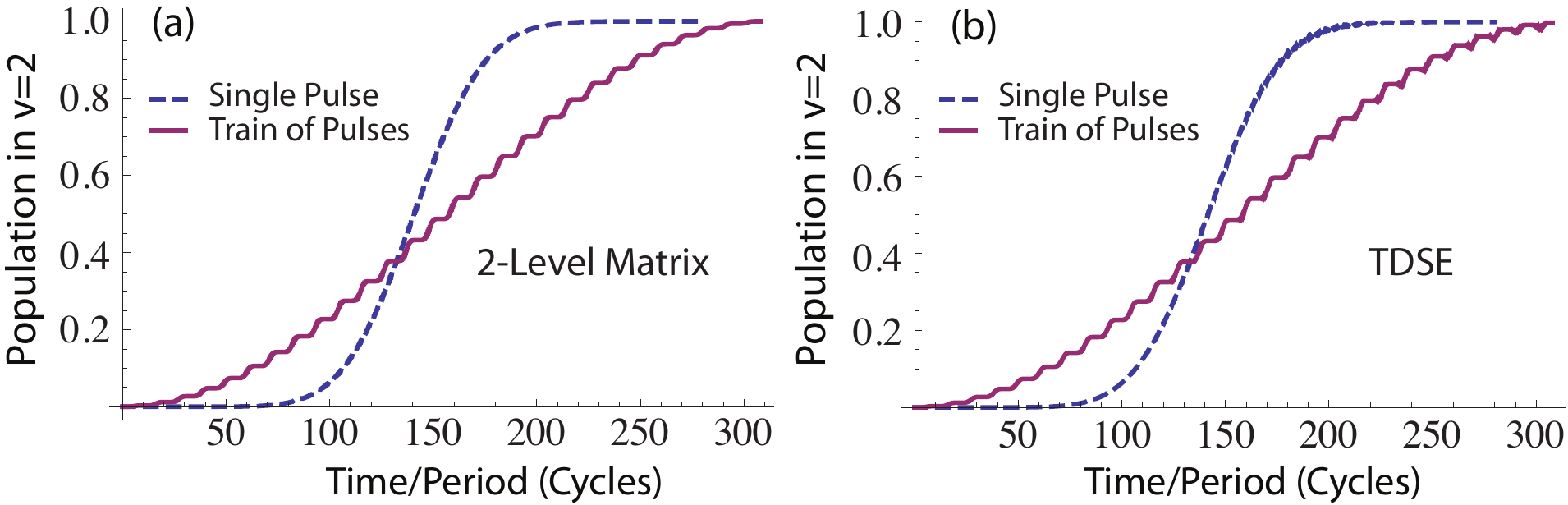}
\caption{
(Color online) Population in the $v=2$ state during the interaction with a chirped laser pulse (280 cycles, dashed line) and a train of 28 chirped pulses (10 cycles each, separated by 1 cycle, solid line),
by solving (a) two-level matrix equation (\ref{2Level_Matrix_VS}) and (b) the TDSE. The peak intensity of each pulse was $10^{13}$ W/cm$^{2}$.
}
\label{fig:ModelvsEffH}
\end{figure}

In this subsection we will test our prediction that the population transfer can be made complete for a certain chirp of the pulse, using both the two-level model as well as full numerical simulations. To this end, we determined the chirp parameters for the transition from the vibrational ground state ($\nu_i=0$) to the second excited state ($\nu_f=2$) as $\mu_{0,0}^{2}-\mu_{2,2}^{2}=(-2.66\pm0.02)$ au and $\mu_{0,2}^{2}=(0.255\pm0.001)$ au, and performed full time-dependent numerical simulations as well as calculations using the two-level approximation for our H$_2^+$ model system. The results of both approaches agree very well (dashed lines in Fig. \ref{fig:ModelvsEffH}). The final probability to find the
system in the excited vibrational state is indeed almost 100\% (exactly, 99.89\%). Since the excitation probability depends on the square area of the pulse envelope, the same result can be achieved by a coherent accumulation of the transition probability using a train of chirped pulses (solid lines in Fig. \ref{fig:ModelvsEffH}). The delay between subsequent pulses must be such that the phase of the initial state at the beginning of a pulse equals the phase at the end of the previous pulse \cite{Peer_Precise_2007}.

\begin{figure}
\centering\includegraphics[width=8cm]{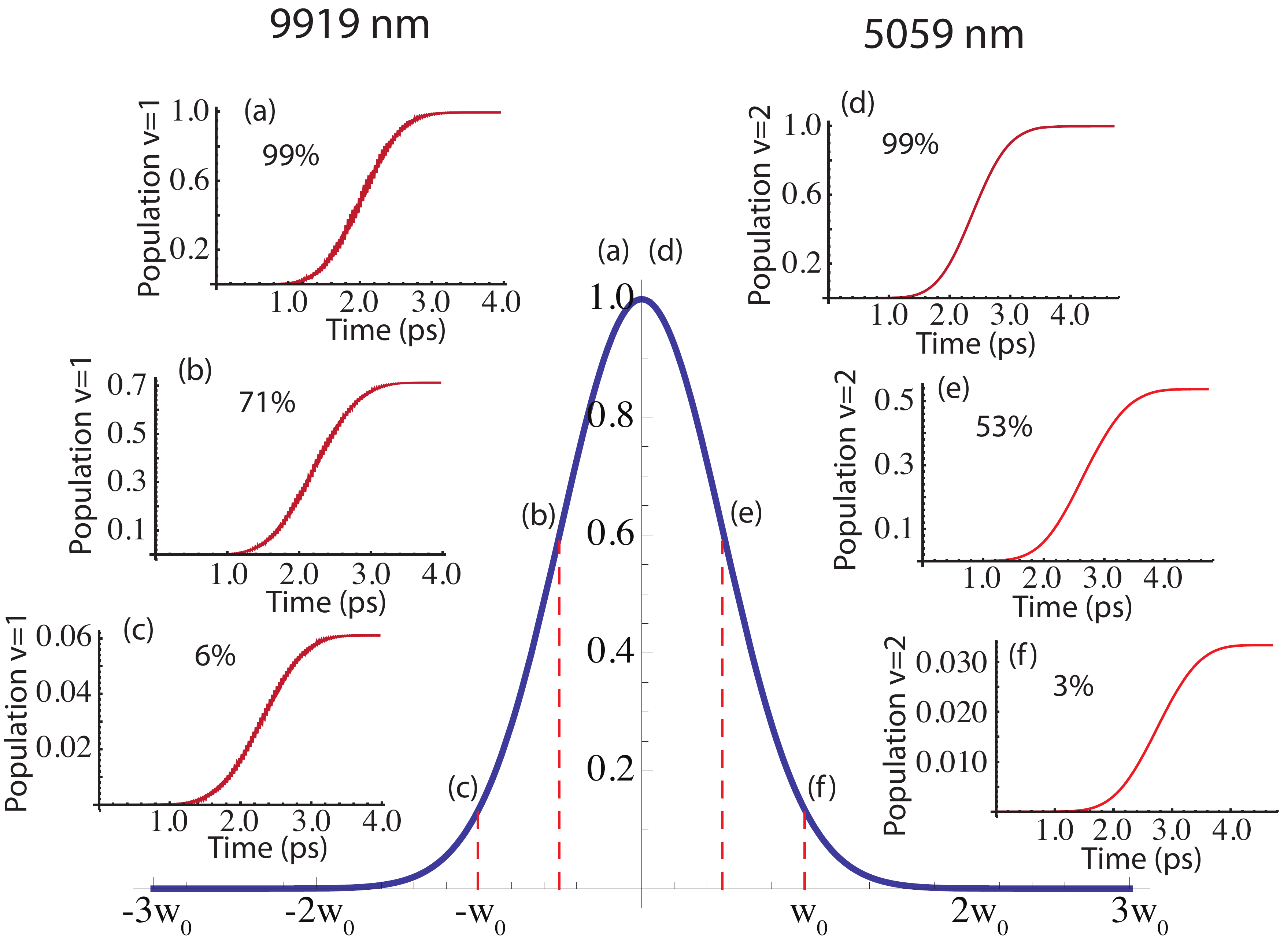}
\caption{
(Color online) Spatial intensity profile of a Gaussian laser pulse (scaled in $10^{13}$ W/cm$^2$).
TDSE simulations for population transfer to (a-c) $\nu = 1$ using a chirped laser pulse with frequency $\omega_{1,0}$ (9919 nm) and 120 cycles and
(d-f) $\nu=2$  using a chirped laser pulse with frequency $\omega_{2,0}$ (5059 nm) and 280 cycles were performed. The results presented in the different panels correspond
to the peak intensities labeled on the intensity profile:
(a, d) $10^{13}$ W/cm$^2$,
(b, e) $6.0\times10^{12}$ W/cm$^{2}$ and
(c, f) $1.3\times10^{12}$ W/cm$^{2}$.
The chirp parameters were determined to maximize the population transfer for the peak intensity at the center of the pulse and kept the same for all simulations. w$_0$ refers to the beam waist.
}
\label{fig:Spatial_Profile}
\end{figure}

Please note that our condition for the chirp parameters depends on the temporal intensity profile, in particular the maximum intensity (c.f. Eq. (\ref{Chirp_Condition})).
In an experiment averaging over the spatial intensity profile (also called focal averaging) cannot be avoided.
This results in a variation of the maximum intensity in the temporal profiles over the focal area. In order to analyze the effect, we have considered a Gaussian (spatial) intensity profile with a
maximum peak intensity of $10^{13}$ W/cm$^2$ (see, Fig. \ref{fig:Spatial_Profile}). We determined the chirp parameters for a complete population transfer from the vibrational ground state to the first (left hand panels in  Fig. \ref{fig:Spatial_Profile})
and second vibrational state (right hand panels in  Fig. \ref{fig:Spatial_Profile}) for the peak intensity at the center and performed TDSE simulations for a set of peak intensities keeping the chirp parameters the same.
As expected, the population transfer at the center is complete but it becomes less and less efficient as the intensity drops towards the wings of the intensity distribution. Still, we find a transfer of more than 73\% (for the transition to the
first excited state) and more than 50\% of the population (for transfer in $\nu=2$) at 60\% of the maximum peak intensity. Even, in the wings of the intensity distribution the transfer is with more of 1\%, not negligible.
A similar reduction in the control efficiency will occur for an unaligned ensemble of molecules, since the effective interaction strength does scale with the cosine of the angle between the polarization direction
of the field and the internuclear axis.

\begin{figure}
\centering\includegraphics[width=8cm]{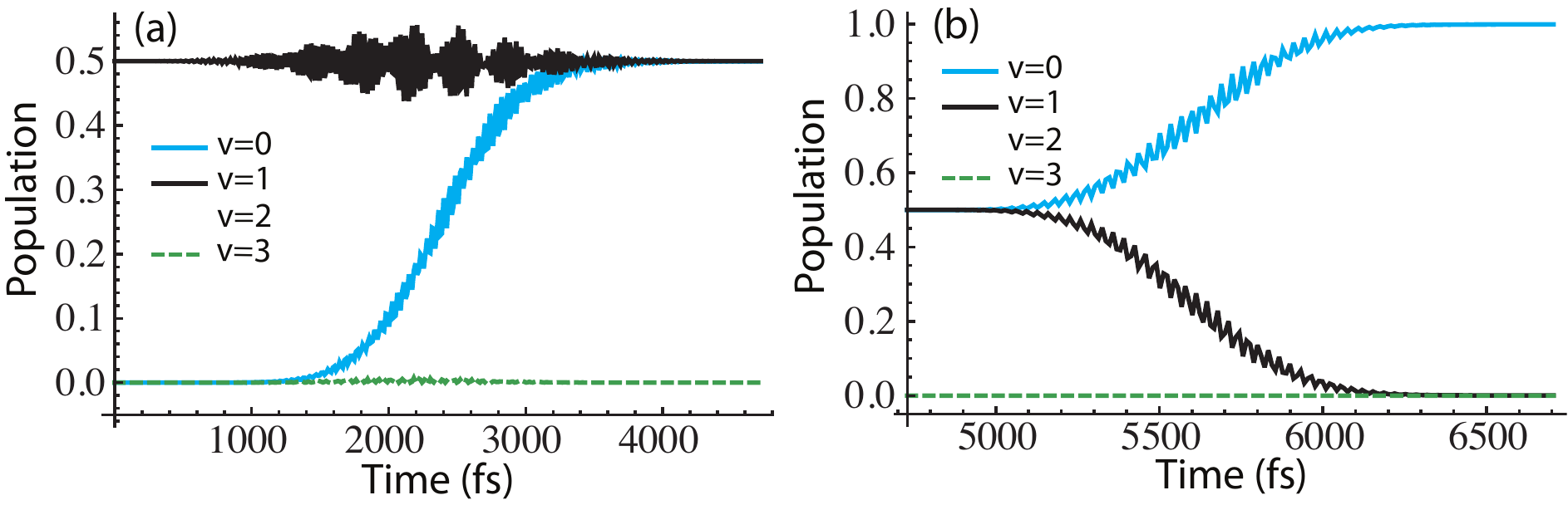}
\caption{
(Color online) Population transfer from a superposition of the excited vibrational states $\nu=1$ and $\nu=2$ to the vibrational ground state.
(a) Population of the lowest vibrational states during the interaction with a first chirped pulse (the same used in Fig. \ref{fig:ModelvsEffH}, $\omega_{2,0} = 9.00\times10^{-3}$).
(b) A second chirped pulse with a peak intensity of $10^{13}$ W/cm$^{2}$, 60 cycles and $\mu_{0,0}^{2}-\mu_{1,1}^{2}=(-1.56\pm0.02)$ au interacts with the remaining superposition after a time delay 10.8 fs.
}
\label{fig:Cooling}
\end{figure}

Finally, we considered population transfer from a superposition of the first and second excited vibrational level ($\nu=1$ and $\nu=2$) to the vibrational ground state. The results of our numerical simulations, shown in Fig. \ref{fig:Cooling}, demonstrate that with two chirped pulses with $\omega_{2,0} = 9.00\times10^{-3}$ au (5059.3 nm) and $\omega_{1,0} = 4.59\times10^{-3}$ au (9919.9 nm) and the appropriate chirps first the population from the second vibrational state (Fig. \ref{fig:Cooling}(a)) and then the remaining population from the first vibrational state (Fig. \ref{fig:Cooling}(b)) is transferred to the ground state.

\section{Conclusions}

In conclusion, we have theoretically and numerically analyzed the coherent control of the population distribution in the vibrational states of the nonpolar hydrogen molecular ion.
Our results unveil that a selective excitation of vibrational states can be achieved via a (net) two-photon transition using infrared laser pulses.
The transitions are accompanied by a dynamical Stark shift, which depends linearly on the laser intensity.
Using a chirped laser pulse the Stark phase shift can be followed and for a given temporal intensity profile a complete coherent transfer can be achieved between two vibrational states of the molecule.
Looking ahead, we anticipate that the same coherent control scheme can be applied to other nonpolar molecules with bound or dissociative electronic states as well as
to the preparation of ro-vibrational states.

\section*{Acknowledgements}

A.P. acknowledges financial support from the Spanish Ministry of Science and Innovation through the Postdoctoral program and the hospitality during his stay at ICFO. J.B. acknowledges support from the Spanish Ministry of Education and Science through its Consolider Program Science (SAUUL - CSD 2007-00013) and through ÒPlan NacionalÓ (FIS2008-06368-C02-01/02). This work was partially supported by the US Department of Energy. We also acknowledge Norio Takemoto, Carlos Trallero-Herrero, and Shaohao Chen for fruitful discussions.


\begin{thebibliography}{99}
%
\bibitem{weinacht99}
T.C. Weinacht, J. Ahn, and P.H. Bucksbaum,
Nature {\bf 397}, 233 (1999).
%
\bibitem{assion98}
A. Assion, T. Baumert, M. Bergt, T. Brixner, B. Kiefer,
V. Seyfried, M. Strehle, and G. Gerber,
Science {\bf 282}, 919 (1998).
%
\bibitem{krems08}
R.V. Krems,
Phys. Chem. Chem. Phys. {\bf 10}, 4079 (2008).
%
\bibitem{koelemeij07}
J.C.J. Koelemeij, B. Roth, A. Wicht, I. Ernsting, and S. Schiller,
\prl {\bf 98}, 173002 (2007).
%
\bibitem{demille02}
D. DeMille,
\prl {\bf 88}, 067901 (2002).
%
\bibitem{viteau08}
M. Viteau, A. Chotia, M. Allegrini, N. Bouloufa, O. Dulieu, D. Comparat, and P. Pillet,
Science {\bf 321}, 232 (2008).
%
\bibitem{ni08}
K.-K. Ni, S. Ospelkaus, M.H.G. de Miranda, A. Pe'er, B. Neyenhuis, J.J. Zirbel, S. Kotochigova,
P.S. Julienne, D.S. Jin, and J. Ye,
Science {\bf 322}, 231 (2008).
%
\bibitem{deiglmayer08}
J. Deiglmayer, A. Grochola, M. Repp, K. M\"ortlbauer, C. Gl\"uck, J. Lange, O. Dulieu,
R. Wester, and M. Weidem\"uller,
\prl {\bf 101}, 133004 (2008).
%
\bibitem{lang08}
F. Lang, K. Winkler, C. Strauss, R. Grimm, and J. Hecker Denschlag,
\prl {\bf 101}, 133005 (2008).
%
\bibitem{ospelkaus10}
S. Ospelkaus, K.-K. Ni, G. Qu\'em\'ener, B. Neyenhuis, D. Wang, M.H.G. de Miranda, J.L. Bohn,
J. Ye, and D.S. Jin,
\prl {\bf 104}, 030402 (2010).
%
\bibitem{danzl10}
J.G. Danzl, M.J. Mark, E. Haller, M. Gustavsson, R. Hart, J. Aldegunde, J.M. Hutson,
and H.-C. N\"agerl,
Nat. Phys. {\bf 6}, 265 (2010).
%
\bibitem{staanum10}
P.F. Staanum, K. H{\o}jbjerre, P.S. Skyt, A.K. Hansen, and M. Drewsen,
Nat. Phys. {\bf 6}, 271 (2010).
%
\bibitem{schneider10}
T. Schneider, B. Roth, H. Duncker, I. Ernsting, and S. Schiller,
Nat. Phys. {\bf 6}, 275 (2010).
%
\bibitem{peters05}
T. Peters, L.P. Yatsenko, and T. Halfmann,
\prl {\bf 95}, 103601 (2005).
%
\bibitem{bergmann98}
K. Bergmann, H. Theuer, and B.W. Shore,
Phys. Rep. {\bf 70}, 1003 (1998).
%
\bibitem{Knight_Laser_1990}
P.L. Knight, M.A. Lauder, and B.J. Dalton,
Phys. Rep. {\bf 190}, 1 (1990).
%
\bibitem{Chelkowski_Raman_1997}
S. Chelkowski and A.D. Bandrauk,
J. Raman Spectrosc. {\bf 28}, 459 (1997).
%
\bibitem{kulander96}
K.C. Kulander, F.H. Mies, and K.J. Schafer,
\pra {\bf 53}, 2562 (1996).
%
\bibitem{Trallero_Coherent_2005}
C. Trallero-Herrero, D. Cardoza and T.C. Weinacht
\pra {\bf 71}, 013423 (2005).
%
\bibitem{Trallero_Strong_2006}
C. Trallero-Herrero, J.L. Cohen, and T.C. Weinacht,
\prl {\bf 96}, 063603 (2006).
%
\bibitem{Eberly_Optical_1987}
L. Allen  and J.H. Eberly,
{\em Optical resonance and two-level atoms}
(Dover Publications, Mineola, 1987).
%
\bibitem{Peer_Precise_2007}
A. Pe'er,  M.A. Shapiro, C.W. Stowe, M. Shapiro, and Jun Ye,
\prl {\bf 98}, 113004 (2007).
%
\end{thebibliography}
\end{document}